\documentclass{article}

\usepackage{arxiv}

\usepackage[utf8]{inputenc} 
\usepackage[T1]{fontenc}    
\usepackage{hyperref}       
\usepackage{url}            
\usepackage{booktabs}       
\usepackage{amsfonts}       
\usepackage{nicefrac}       
\usepackage{microtype}      
\usepackage{lipsum}
\usepackage{graphicx}
\usepackage{amsmath}
\usepackage[font=normalsize]{caption}
\usepackage{graphicx}
\usepackage{float}
\usepackage{subcaption}
\usepackage[labelfont=bf]{caption}
\usepackage{caption}
\usepackage{authblk}
\graphicspath{ {./images/} }
\title{An Investigation into the Causal Mechanism of Political Opinion Dynamics: A Model of Hierarchical Coarse-Graining with Community-Bounded Social Influence
}

\author[1,*]{Valeria Widler}
\author[2]{Barbara Kamińska}
\author[3]{André C. R. Martins}
\author[4]{Ivan Puga-Gonzalez}

\affil[1]{Modeling and Simulation of Complex Processes, Zuse Institute Berlin, Berlin, Germany}
\affil[2]{Faculty of Management, Wroclaw University of Science and Technology, Wrocław, Poland}
\affil[3]{Escola de Artes, Ciências e Humanidades, Universidade de São Paulo (EACH-USP), São Paulo, Brazil}
\affil[4]{Center for Modelling Social Systems, Norwegian Research Center (NORCE), Kristiansand, Norway}
\affil[*]{Corresponding author E-mail: valeria.widler@gmx.de}

\begin{document}

\maketitle
\begin{abstract}
The increasing polarization in democratic societies is an emergent outcome of political opinion dynamics. Yet, the fundamental mechanisms behind the formation of political opinions, from individual beliefs to collective consensus, remain unknown. Understanding that a causal mechanism must account for both bottom-up and top-down influences, we conceptualize political opinion dynamics as hierarchical coarse-graining, where microscale opinions integrate into a macro-scale state variable. Using the CODA (Continuous Opinions Discrete Actions) model, we simulate Bayesian opinion updating, social identity-based information integration, and migration between social identity groups to represent higher-level connectivity. This results in coarse-graining across micro, meso, and macro levels. 
Our findings show that higher-level connectivity shapes information integration, yielding three regimes: independent (disconnected, local convergence), parallel (fast, global convergence), and iterative (slow, stepwise convergence). In the iterative regime, low connectivity fosters transient diversity, indicating an informed consensus. In all regimes, time-scale separation leads to downward causation, where agents converge on the aggregate majority choice, driving consensus. Critically, any degree of coherent higher-level information integration can overcome misalignment via global downward causation. The results highlight how emergent properties of the causal mechanism, such as downward causation, are essential for consensus and may inform more precise investigations into polarized political discourse.  
\end{abstract}


\section{Introduction}
Political discourse in many established democracies is deteriorating \cite{Borbarth2023}. Conflicts over immigration, cultural issues and international cooperation have superseded the classic left-right economic cleavage, consolidating what is often referred to as a “cultural cleavage” \cite{Borbarth2023,Dalton2022,Enke2023}. This restructuring of the issue landscape is accompanied by a problematic transformation of political discourse into a polarized contest between two collective identities on either side of the new cleavage \cite{Borbarth2023}, i.e., affective polarization. 
A “healthy” \cite{Jost2022, tornberg2022} political discourse is shaped by a discussion of diverse points of view that combine to form consensus. Conversely, affective polarization occurs when two opinion groups are unwilling to listen to each other’s perspectives \cite{Hobolt2021}, have strong regard for and trust in the opinion ingroup and dislike and distrust for the outgroup \cite{Iyengar2012}. In this case, individuals align strongly with their own groups while disregarding perspectives from others, impeding integrative discourse and consensus among groups on an increasing array of issues \cite{Iyengar2015}.

Research into the causes for increasing dysfunction in political discourse often centers around changes to the communication architecture induced by the internet \cite{Barbera2020, Bohmelt2020, Ohme2020}. The mechanism underlying political discourse that is affected by these changes in network connectivity is however not comprehensively known. We propose to investigate the bidirectional causal mechanisms underlying information integration in political opinion dynamics and their role in the emergence of collective consensus through an agent-based model. We conceptualize public discourse as a collective sense-making process that aggregates inferences of individuals into a unitary perception by way of hierarchical coarse-graining. By implementing likely bottom-up mechanisms of human collectives in an agent-based model, we seek to investigate the emergence of the top-down mechanism of downward causation. We then go on to test how higher-level connectivity impacts the dynamics and the achievement of a consensus.

How can we model public discourse? Human societies are embedded in environments of immense complexity. Survival depends on the computation of predictions that reduce the uncertainty in the environment, informing collective action and a unitary perception of the world. Flack \cite{Flack2017a} coined the term ‘endogenous coarse-graining’ for this process of collective sense-making in natural systems. Generally, coarse-graining is the lowering of the resolution of a system's information across scales at minimal loss of information. A system's microscopic information is compressed or smoothed over and abstracted, but along a functional dimension. It serves as an effective way to understand a system for the purpose of prediction and intervention \cite{Liao2019}. 
We take the perspective that human societies employ endogenous coarse-graining to construct a collective perception of the environment. Consequently, political opinion formation is a collective computation where individuals process information about the environment (accumulation phase) and these inferences are aggregated into a collective perception through opinion exchanges within and across local communities (consensus phase) \cite{Flack2017a}. We are interested in the consensus mechanism that regulates the hierarchical coarse-graining of opinions across scales via upward and downward causation into a unitary joint opinion \cite{Harte2024}.

Since components in nature integrate information on the basis of heuristics, the simplification is an approximation and not a direct translation of the environmental inferences \cite{Flack2017a}. Subsequently, how the mechanism clusters information across scales, can be beneficial to the collective estimation or a source of bias \cite{Flack2017a, Smaldino2017}. Research in epistemological studies on decision-making in heterogeneous groups suggests that a transient diversity of opinions, an intermediate rise in diversity of opinions before convergence, indicates a more informed consensus, whereas herding (i.e. a quick convergence to a global consensus) signifies a time efficient, however less informed consensus \cite{Smaldino2022}.

At the microlevel, where bottom-up mechanisms drive information integration, social learning strategies (SLS), such as preferential integration of social information from individuals sharing the same social identity, modulate the flow of information \cite{Abrams1990, Smaldino2017, Hamalainen2023, Laland2004, Hawkins2023}, acting as a gatekeeping criterion for information integration based on similarity \cite{Steiglechner2024}. When shared local identities act as the social influence heuristic, the society is essentially partitioned into ‘epistemic communities’ \cite{Albarracin2022} wherein information is integrated freely, leading to an alignment of beliefs, preferences, and behaviors. Between these communities information integration is impeded, since based on this mechanism alone no opinions should be integrated by a sender from a different epistemic community. Higher-level connectivity is thus facilitated by other means. One possibility are types of noise that act on the social interaction at different points, e.g. ‘selectivity noise’ \cite{Steiglechner2024}, where social information that a social learning strategy would typically ignore is included nonetheless every so often. 

We model how people simplify and integrate information using Bayesian inference within the Continuous Opinions and Discrete Actions (CODA) model \cite{Martins2008}. In this model, individuals hold internal beliefs about different choices regarding a specific issue and express the one they consider most likely to be correct, i.e., its preferred choice. They share this preferred choice with others in one-on-one exchanges. The extent to which someone adjusts their beliefs in response to another’s preferred choice depends on whether they belong to the same epistemic community (represented by a label). If they share a label, they integrate information, if they are from different groups, the influence varies—ranging from being ignored to slightly affecting the receiver’s belief. When agents receive enough input favoring a different choice, their preferred choice changes. They may also switch communities by dropping their current label and adopting a new one, allowing preferred choices to spread between groups. By adjusting how often agents switch communities, we can observe how different levels of cross-community connectivity influence the coarse-graining process, shaping how individual preferred choices evolve from the micro level into collective patterns at the macro level.

Using this model, we investigate our research questions: What is the causal mechanism, including both upward and downward causation, that leads to consensus? How does higher-level connectivity affect this mechanism? How does the mechanism relate to polarization? To address these questions, we measure whether preferred choices among agents and communities align or diverge over time. Alignment occurs when all components at a given scale increasingly share the same preferred choice, and global consensus is reached when there is local as well as global alignment. 

We systematically varied all model parameters within specified ranges to evaluate their influence on the dynamic outcomes of information integration (e.g., local and global convergence, diversity indices). We found that the probability of dropping the label (PDL) is the dominant factor determining system dynamics and regime transitions, while other parameters, such as the number of labels, issues, choices, ‘ignoring on/off’, strength-of-influence, and multi-issue-discourse, primarily modulate the critical PDL thresholds. Our findings reveal that varying levels of inter-group connectivity (PDL) give rise to three distinct information integration regimes:

\begin{itemize}
\item Independent Information Integration Regime: When inter-group connectivity (migration) is absent, agents reach local alignment, but no global information aggregation occurs. Information is coarse-grained only at the local level.

\item Parallel Information Integration Regime: At high inter-group connectivity, local and global alignment happen simultaneously, leading to rapid global consensus (local and global alignment) in a single phase. This corresponds to herding—a fast but less informed consensus process \cite{Smaldino2017}.

\item Iterative Information Integration Regime: At low levels of inter-group connectivity, information integration occurs in three phases (I:local alignment, II: global alignment, III: local alignment), with slow but informed convergence that gives rise to a transient diversity\cite{Smaldino2022}. 

\end{itemize}

These results highlight how simple social heuristics can interact with network dynamics to produce complex, multi-level political phenomena. The discovered bidirectional causal mechanism suggests that polarization may arise from a disturbance in higher-level information integration that disrupts the emergence of downward causation necessary to reach consensus. Further development of this model could offer new perspectives on the forces that contribute to democratic erosion, the reinforcement of ideological bubbles, and strategies to encourage more functional public discourse.  

\section{Background}
Collective decisions in democratic societies emerge from the aggregation of individual judgments regarding which actions governments should prioritize to promote social welfare \cite{Lorenz2018}. Like adaptive biological systems, human societies consist of semi-independent individuals, whose interactions result in ordered behavior at aggregate higher levels \cite{Flack2017b}. In political discourse, this manifests as consensus emerging from diverse perspectives on societal issues. As a complex adaptive system, the mechanism underlying democratic discourse likely exhibits bidirectional cross-scale causation where micro- and macro-scale dynamics are ‘entwined’ \cite{Harte2024}. Understanding these dynamics requires considering both bottom-up mechanisms and top-down causal influences, yet “analysis of the dynamics of multilevel systems rarely attempts to capture downward causation” \cite{Harte2024}. 

The theory of collective computation \cite{Flack2017a, Flack2017b, Daniels2016, Brush2013} proposes that individuals act as information processors, coarse-graining environmental information into models during an accumulation phase \cite{Flack2017a}. In a second, the consensus phase, individual models are hierarchically coarse-grained into a collective perception, a process termed ‘endogenous coarse-graining’ \cite{Flack2017a}. This coarse-grained process reduces uncertainty over fast microscopic behavior, producing a lower-resolution variable of the state of the system \cite{Daniels2016, Harte2024}. When coarse-grained models stabilize across individuals, a coherent aggregate scale emerges, allowing a slower state variable to form. This time-scale separation of the micro- and the higher-level serves as a stable ‘social environment’, enabling individuals to tune their behaviors to collective patterns \cite{Flack2017a}. As a result, the coarse-grained aggregate exerts downward causation, reducing entropy and creating order, ultimately leading to convergence. 

Downward causation suggests that higher-level components (e.g., communities), can constrain the future of all lower-level components (e.g., individuals) \cite{Flack2017a, Santos2024}. Tuning occurs when individuals identify regularities in coarse-grained variables and adapt their estimates accordingly. When tuning is partial or imprecise, agent estimates diverge, a phenomenon called Apparent Downward Causation or weak emergence \cite{Flack2017a}. In contrast, Effective Downward Causation or strong emergence occurs when individuals’ estimates cohere, leading to perfect tuning and convergence in the sampling environment \cite{Flack2017a}.

Effective downward causation occurs when:
\begin{itemize}
\item aggregate properties predict the future state of lower level components in the system
\item aggregate properties are stable and exhibit time-scale separation from the level below
\item all components tune to the same aggregate properties 
\item increasing convergence of components’ estimates of aggregate properties
\item convergence comes with an increase in mutual information between the level exerting downward causation and the level below 
\end{itemize} \cite{Flack2017a}

Using this framework of “coarse-graining as a downward causation mechanism” \cite{Flack2017a}, we can quantify top-down causal influence in our model to analyze its role in system dynamics. For the remainder of the paper we use ‘downward causation’ to refer specifically to ‘effective downward causation’.

To understand the bottom-up mechanism, we need to understand how human groups implement endogenous coarse-graining. First, detailed information is coarse-grained at the microlevel. In a hierarchically coarse-graining system, this coarse-grained information is further abstracted into aggregate variables across scales, forming a hierarchy where components at one level are grouped into higher-level state variables, simplifying the description of the system as we go up the scale. Upward causation occurs when interactions at the lower lead to changes in higher-level components \cite{Mediano2022}. In the context of opinion dynamics, upward causation describes how the information integration mechanism at the microlevel influences the coarse-grained state variable at the higher level, in this case, the distribution of opinions in a group.

Coarse-graining is advantageous because the aggregation of microlevel data into the macro level representation, with what Flack terms “slow variable construction” \cite{Flack2017a}, can detect subtle or slow-changing patterns in the environment. In contrast, rapid aggregation of information (fast variable construction) may smooth over microdata too quickly, risking loss of relevant information. Coarse-graining improves the signal-to-noise ratio, offering a time- and energy-efficient way to build an effective theory of the world that allows prediction and intervention without requiring detailed knowledge of causal origins \cite{Flack2017a}. Another benefit is the reduction of uncertainty, as the system converges on a single coarse-grained model. How noise is smoothed across and within scales affects how faithfully the macro-level reflects the original system. In natural systems, this smoothing is often guided by heuristics that prioritize efficiency over accuracy \cite{Flack2017a}. To define a bottom-up mechanism for political opinion dynamics, we explore simple heuristic mechanisms that might underlie this coarse-graining.

For the microlevel coarse-graining mechanism, we adopt a Bayesian inference perspective, where beliefs are represented as probabilistic states. We use a variant of the Continuous Opinions Discrete Actions (CODA) model \cite{Martins2008}, where choices represent the possible hypotheses on an issue, and their input frequencies, encode expected probabilities. This forms an agent’s choice distribution for a given issue, informing it on the plausibility of a choice. The relative entropy of the distribution reflects uncertainty: high entropy indicates maximum uncertainty (equal evidence across choices), while low entropy reflects certainty and convergence on a single choice. The ‘preferred choice’ is the one with the highest frequency and this is the agent’s current opinion that is communicated to others. An expressed opinion is coarse-grained, as it condenses the agent’s past social interactions into a single output. The recipient does not access the full interaction history—rather, they infer the sender's position from the expressed opinion alone.

Another key element of the micromechanism is how agents select sources of information for incorporation into their internal model of the world. While people differ along many dimensions, such as gender, age, or socioeconomic situation, social identity theory and self-categorization theory \cite{Tajfel1979, Abrams1990} suggest that individuals rely on a subjective social frame of reference dividing the social world into social categories of similarity and difference. This process results in in- and outgroups shaping patterns of social influence, ingroup conformity and outgroup differentiation \cite{Abrams1990, Smaldino2025}. While there are other heuristics, like authority or expertise, that can activate social influence, a person’s social identity remains a primary social frame of reference and has been shown to play a key role in political opinion formation \cite{Abrams1990}. Ingroups can be viewed as ‘epistemic communities’, a local sense-making environment where information is trusted and shared, creating shared mental models, preferences and codes of conduct \cite{Albarracin2022}. These communities create relatively stable social environments \cite{Flack2017a} and reflect the multi-scale nature of human social organization, where strong local interactions give way to weaker ones across increasing social distance, from kin to stranger \cite{RamosFernandez2020}. Additionally, affective polarization has been linked to social identity processes, particularly those driven by perceived inter-group dynamics \cite{Barbera2020, You2024}.

To reflect this self-categorization into ingroups of social influence, we extend the CODA model by introducing ‘labels’—identity markers that determine whether an interaction leads to information integration. To enable information sharing across communities, we implement noise into the social influence heuristic \cite{Steiglechner2024} via a probability of migrating between labels, representing the assimilation of individuals into another epistemic community. 

In this model, we implemented two bottom-up mechanisms: the shift in preferred choice of an agent and the probability of agents switching their label. From these two sources of change emerge the regimes and phases observed in the simulations. The coarse-grained output, the preferred choice, becomes the input for the higher scale (Fig. 1). Due to the hierarchical clustering of components in our model, a change in preferred choice of an agent shifts the composition of the preferred choice distribution within the label which may in turn accumulate and produce a shift in the label’s preferred choice and thus a change in the global distribution in a hierarchical chain of upward causation. 

\begin{figure} 
    \centering
    \includegraphics[width=1\textwidth]{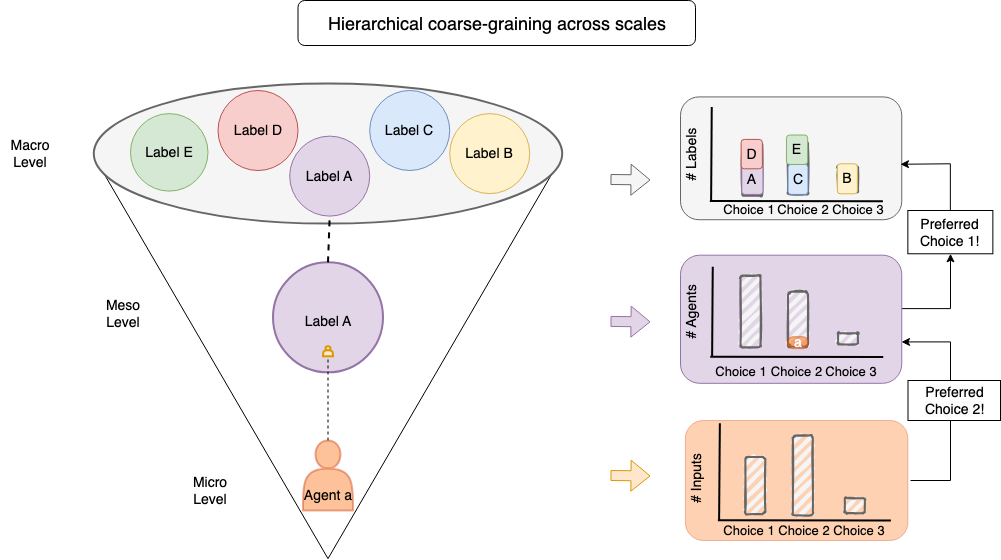}
    \caption{The hierarchical coarse-graining of preferred choices from agent to global level is shown on the left, while the process of upward causation in the hierarchical coarse-graining scheme is shown on the right}
    \label{fig:1}
\end{figure}

The shift in preferred choice, i.e. upward causation, occurs when the amount of evidence favoring a divergent choice overtakes the current one. We term this threshold the activation barrier (AB). While our notion of the activation barrier (AB) arises from the logic of our model, it is conceptually anchored in the energy barrier concept in complex systems. In statistical physics and nonlinear dynamical systems, transitions between stable states require overcoming an energy barrier, which determines the likelihood of state change \cite{Strogatz2001}. Similarly, here, the AB represents the cumulative divergent social input required for an agent to change its preferred choice. A larger activation barrier requires a greater accumulation of social inputs, which in turn means an increase in the number of interactions, and therefore timesteps, required before an agent can change their preferred choice. 

\begin{figure} 
    \centering
    \includegraphics[width=0.8\textwidth]{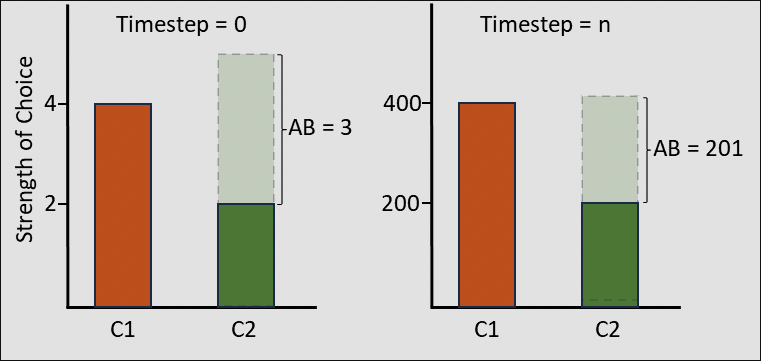}
    \caption{Activation barrier (AB) 
At initialization, the agent’s internal choice distribution consists of six units of information (left panel), the AB is low. As time goes by, the agent integrates information and the AB becomes higher. 
C1-C2 represent choice 1 and 2 respectively}
    \label{fig:2}
\end{figure}

In our model, the AB is dynamic and increases as a function of time: the more information is stored in the agent’s choice distribution, the larger the sum of inputs needed to overcome the AB due to the fixed size of the input over the simulation run, reducing its impact on the agent’s distribution. Fig. 2 illustrates how the AB increases from 3 at t = 0 to 201 at t = n. We refer to the sum of inputs between preferred choice changes as the input term. A second aspect of the activation barrier is linked to relative entropy. Agents with strong evidence for one choice and little for others are certain, making them more resistant to change. In contrast, agents with more evenly distributed choice strengths (high relative entropy) have higher uncertainty and lower activation barriers. The AB at the local community level is given by the activation barrier of agents in that community, i.e. their level of stubbornness, and the relative entropy of the distribution. The global AB is defined by the label’s AB and its level of relative entropy, with high relative entropy signifying a high AB and v.v.

We hypothesize that when agents’ internal distributions are increasing in information content, and when they are increasing in certainty, the AB rises, lengthening the time required for preferred choice change to occur. This stabilization reinforces time-scale separation, making higher-level distributions progressively more resistant to changes at lower levels. These ideas in combination with Flack’s indicators for downward causation form the basis for our analysis of the emergent causal mechanism.

Since this paper focuses on the consensus phase of collective computation, we concentrate on tuning, the process by which agents estimate regularities in the social environment, i.e. the coarse-grained variable \cite{Flack2017a}. Here, the social environment is defined as the distribution of choices within the agent’s local community. At any point during the simulation, an agent’s model represents its current estimation of that distribution, shaped by past interactions. To isolate the intrinsic mechanism of information integration, we start the simulation with a uniform distribution of choices. This maximum entropy condition minimizes the influence of prior information, allowing the dynamics to emerge solely from the model’s structure and interactions \cite{Yuan2024}. 

By implementing a bottom-up mechanism at the microlevel and tracking the hierarchical transmission of information via upward causation, our goal is to identify the emergent downward causal influence of the collective computation using Flack’s \cite{Flack2017a} indicators for downward causation outlined in this section. Because the terminology for describing the evolution of distributions is sparse, we define key concepts for clarity. Alignment is defined as the convergence on a single preferred choice within a given scale—for example, when all agents within a label share the same majority choice (local alignment) or when all labels converge on the same preferred choice (global alignment). Misalignment, conversely, refers to the persistence of diverse preferred choices within a distribution. Global consensus is achieved only when both local and global alignments are realized, meaning all agents across all labels share the same choice.

In summary, our theoretical framework suggests that opinion formation and consensus-building are driven by hierarchical coarse-graining mechanisms, where micro level interactions shape local and global distributions through upward causation, and after stability and time-scale separation emerge, the respective higher-level distribution exerts downward causation on the level below. Understanding how different levels of connectivity influence these dynamics is essential to understand political discourse higher-level outcomes such as polarization. To systematically explore these mechanisms, we implement an agent-based model that captures the interplay between individual learning, epistemic communities, and hierarchical opinion integration. The following section outlines the structure of our model, its key parameters, and the methodological approach used to analyze the emergent dynamics.

\section{Methods}

\subsection{Model Overview}

The model is written in Python v 3.11.5. Here we present a brief description, a full ODD+D protocol can be found as Supporting Information \cite{Kaminska2025}.

The model simulates an artificial society inhabited by 2500 human agents who have a set of independent issues, varying from 2 to 10. Each issue represents a specific problem or topic the agents face, and for each issue, agents evaluate a set of choices, varying from 2 to 10. Agent holds an internal belief about which choice is the best solution for a given issue i.e., the preferred choice. This belief is represented by a vector of choice strengths (Fig. 3), where each value corresponds to the agent's estimation of how strongly each choice is the best solution. The preferred choice is the one with the highest choice strength, which is communicated to others during interactions (Fig. 3).

\begin{figure} 
    \centering
    \includegraphics[width=0.5\textwidth]{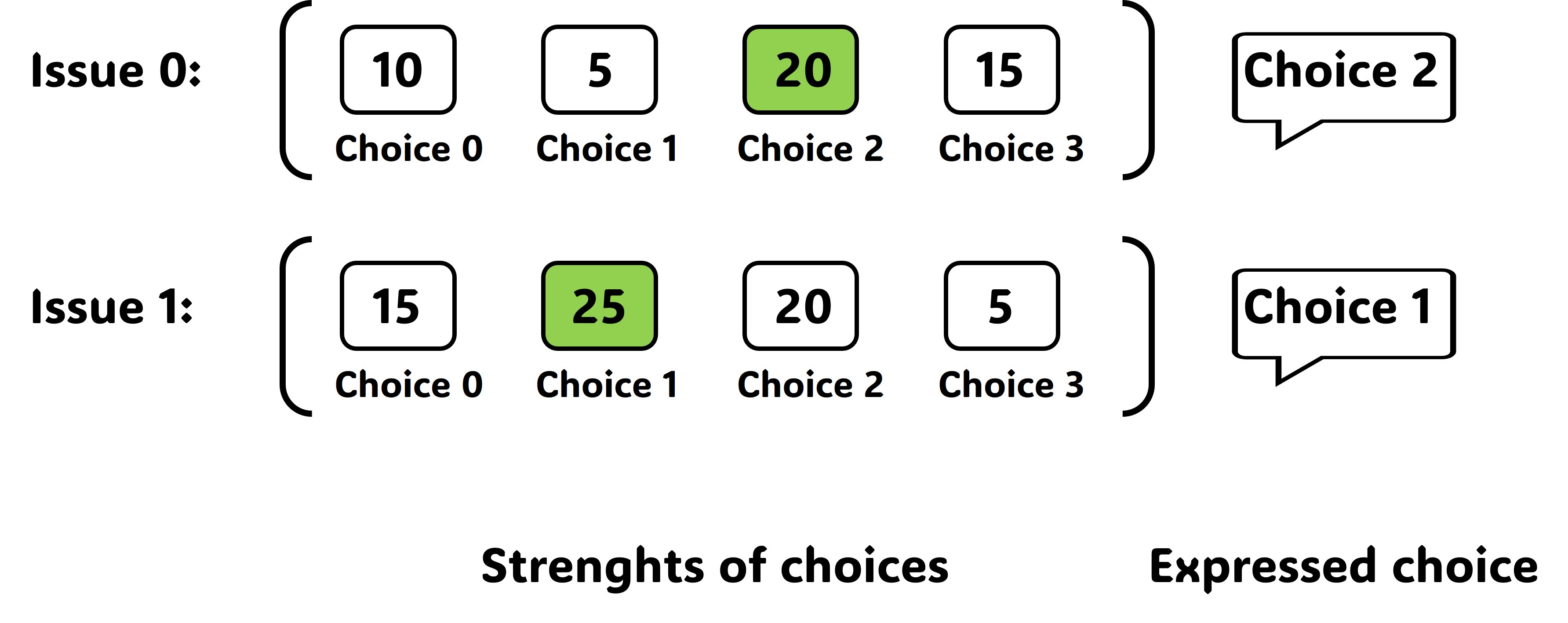}
    \caption{Graphical representation of agents' choices on issues  
Agents evaluate the strength of each choice and express the one with the highest strength }
    \label{fig:3}
\end{figure}

 To simplify the model, choices’ strengths are updated using a Bayesian inference process. To reduce computational complexity, the probabilities are transformed into the log-odds domain. This change allows the use of a simple additive rule, enabling efficient calculation of the strength of each choice while preserving probabilistic reasoning. Agents thus hold a numerical vector per issue, where values represent the strength of the choice. Further, agents also belong to a label. The label value ranges from none (no label) to 1, …, \textit{number\_of\_labels} in the system. Labels influence the frequency and trustworthiness of interactions: agents trust and interact more frequently with those sharing the same label.  Every time step, agents may drop their current label with probability \textit{prob-dropping-label} or adopt a new one with probability \textit{prob-adopting-a-label} (Fig. 4). 

\begin{figure} 
    \centering
    \includegraphics[width=1\textwidth]{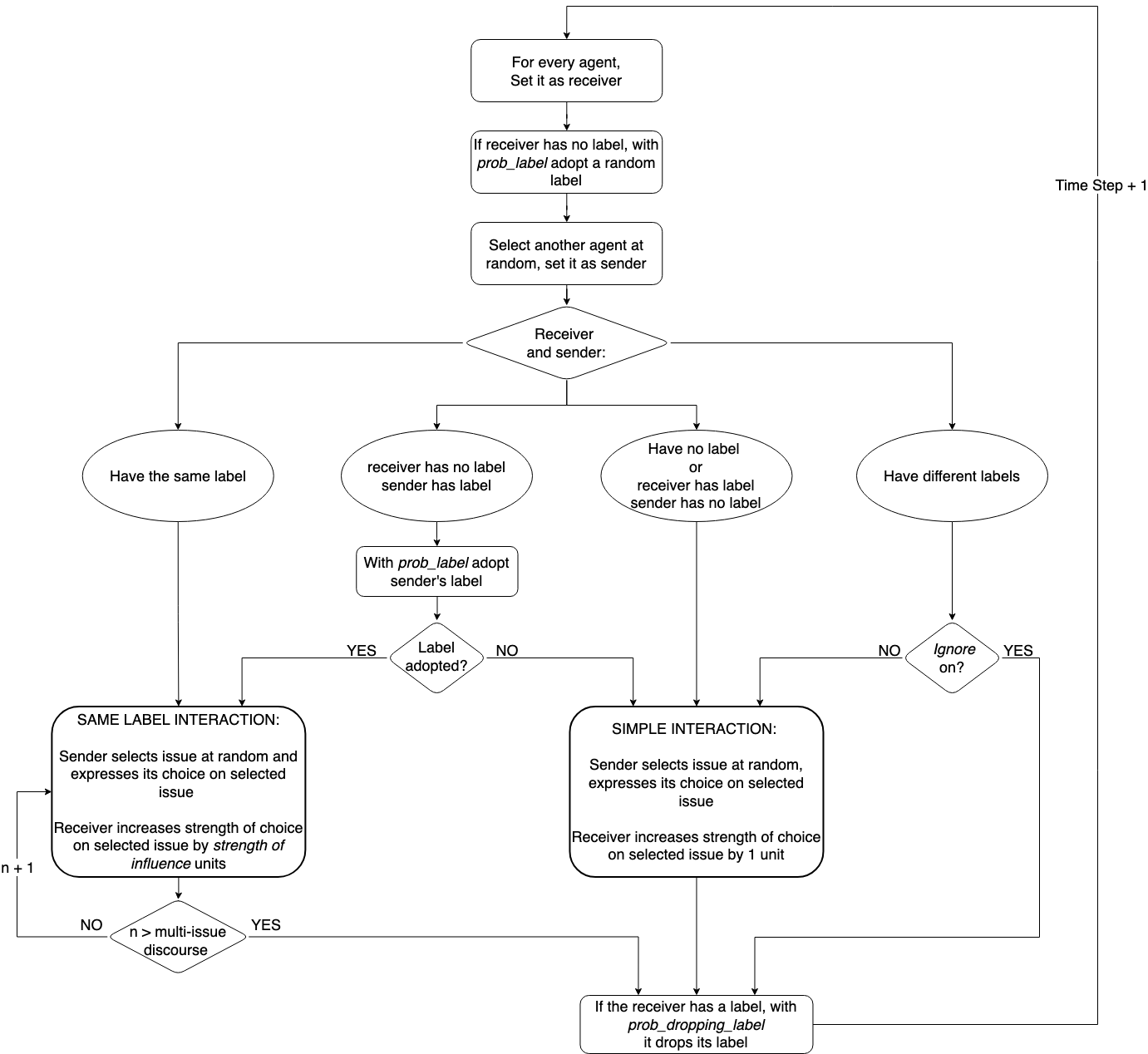}
    \caption{Flow diagram representing algorithm of the interactions between agents. Note that whether Ignoring is ON or OFF is a parameter of simulation (fixed for given run)}
    \label{fig:4}
\end{figure}

At initialization the parameters are set as outlined in Table 1. The rationale for these choices is as follows: Probability of dropping label (\textit{prob-dropping-label}): values higher than 0.01 lead to rapid consensus across labels. We explored the 0 to 0.005 range as it preserves diverse dynamics. Although these values seem small, they result in a steady flow of individuals changing labels every time step (e.g., at 0.005, \textasciitilde12 individuals switch labels per time step in a population of 2500). The range for number of labels and issues was chosen to ensure manageable, interpretable outcomes and computational efficiency. Higher values affect the speed of the dynamics but do not yield qualitative differences. Number of choices per issue: A maximum of 10 choices was selected to approximate the cognitive limit for an average individual. Multi-issue discourse and strength of influence: These parameters represent minimum, medium, and extreme influence levels and were set arbitrarily to explore diverse scenarios. Probability of adopting a label: Fixed to simplify the exploration of other variables, as varying it would primarily affect the speed of dynamics rather than qualitative outcomes. Population size: Fixed at 2500 for computational efficiency.  

The simulation starts with agents randomly assigning low initial choice strengths draw from a uniform distribution [0,3]) for all choices and selecting a random label. Thereafter, the simulation starts and at each time step, individuals hold a dyadic interaction where they exchange preferred choices on multiple issues and update their beliefs accordingly (Fig. 4).

\begin{table}[ht]
    \centering
    \renewcommand{\arraystretch}{1.2} 
    \begin{tabular}{|p{3cm}|p{4cm}|p{7cm}|} 
        \hline
        \textbf{Parameter} & \textbf{Values explored} & \textbf{Description} \\ 
        \hline
        Pop-size & 2500 & Number of agents in the model \\ 
        \hline
        num-labels & [2,6,10] & Number of labels in the system \\ 
        \hline
        num-issues & [1,5,10] & Number of issues (problems or topics) that agents face \\ 
        \hline
        num-choices & [2,5,10] & Number of potential choices (solutions) for each issue \\ 
        \hline
        multi-issue discourse & [1,5,20] & Number of issues discussed during a single interaction with same-label agents \\ 
        \hline
        strength-of-influence & [1,5,20] & Degree of influence that agents with the same label exert on each other \\ 
        \hline
        Ignoring & [on/off] & Whether agents ignore the choices expressed by agents from a different label \\ 
        \hline
        prob-dropping-label & [0.000, 0.0001, 0.0005, 0.0010, 0.0050, 0.0100] & Probability that an agent drops its current label \\ 
        \hline
        prob-adopting-a-label & 0.1 & Probability that an agent without a label adopts a new label \\ 
        \hline
    \end{tabular}
    \caption{Model parameters and their explored values.}
    \label{tab:parameters}
\end{table}

\subsection{Interactions}

Every time step, each agent selects another at random to interact with. Thus, every agent can always potentially interact with every other in the model. In this sense, the model represents a fully connected network. Once an interaction partner is selected, the active agent becomes the receiver and the randomly selected partner the sender. The sender picks an issue at random and communicates its preferred choice for that issue to the receiver. The preferred choice is the one with the highest strength. If two or more choices have the same strength, one of these is selected at random. Thus, while the strength of choices is represented as continuous, the receiver only observes the sender’s preferred choice (Fig. 3).

The outcomes of these Interactions differ depending on the labels held by the sender and receiver. There are four possible combinations of sender-receiver labels, grouped into two interaction types: simple interactions and same-label interactions. For clarity, the interaction process is summarized in a flow diagram (Fig. 4).   

\textit{Simple interactions}. Simple interactions occur when the sender and receiver have no label, only one of them has a label, or both have different labels but the receiver does not ignore the sender (\textit{ignoring} is off, Fig. 4). In these interactions, the sender communicates its preferred choice and the receiver updates the strength associated with observed choice by adding +1. For example, if a sender, holding a different label than the receiver, communicates choice B as its preferred choice on an issue, the receiver increases the strength of choice B by +1, provided that \textit{ignoring} is off. Conversely, if \textit{ignoring} is on, the receiver disregards the sender’s input.

\textit{Same-label interactions}. In interactions when both agents share the same label, the sender has greater influence on the receiver. The \textit{strength-of-influence} parameter determines the degree of this influence. Upon observing the sender’s preferred choice, the receiver updates its internal representation of the issue by increasing the strength associated with the observe choice by \textit{strength-of-influence} units. Further, the \textit{multi-issue discourse} parameter determines the number of issues discussed during the interaction. Consequently, the sender may express its preferred choice for one or more issues during each interaction (Fig. 4). For example, if \textit{multi-issue discourse} is set to 3, the sender shares its preferred choices for three randomly selected issues, and the receiver updates the strengths for all three according to \textit{strength-of-influence} (Fig. 4).

\subsection{Simulations, Data Collection and Analyses}

We explored the parameter space outlined in Table 1. For each parameter combination, we ran 10 simulations. Here, we present results where multi-issue discourse and strength-of-influence were fixed at 20, while prob-dropping-label was restricted to 0, 0.0001, and 0.01. This selection was made to illustrate the model’s key dynamics more clearly, as these parameters primarily influence the speed at which the system reaches stability but do not introduce qualitative differences in the model's dynamics. For a detailed presentation of all parameter combinations, see the Supporting Information (Kamińska, 2025). 

During the simulations every 200 time steps we collected for all agents’ labels and current opinion distribution. With the data collected, we calculate two indexes that help us describe the state of the system: the Cosine index and the Shannon-Wiener (SW) diversity index. These two indexes focus on two key aspects of the system: the alignment of choices within groups sharing the same label (SW index) and the diversity of choices across labels (Cosine index). 
Both, the SW diversity index and the Simpson index, are valid measures of diversity within a given community. The fact that there is a logarithm in the SW index formula makes it more sensitive to minorities. In the case of our model it allows us to more accurately determine when the unanimity among agents holding the same label is broken (when there are agents expressing choices different than the majority).

The SW index assesses how similar the choices are among agents holding the same label. For each label and each issue, we take the frequencies of choices and calculate the SW diversity index, which is basically the entropy normalized by the logarithm of the number of possible choices, so it ranges between [0, 1]. A value of 0 indicates unanimous agreement within a label on a given issue (label alignment), while a value of 1 suggests a uniform distribution of choices (label misalignment). For the sake of clarity in our visual representations, we present the SW diversity index averaged over issues and labels. The mathematical formula to calculate the SW index is presented in equation 1: 
\begin{equation}
    \textit{Shannon - Wiener Index} = \frac{1}{n_{\text{labels}} \cdot n_{\text{issues}}} 
    \sum_{l=1}^{n_{\text{labels}}} \sum_{i=1}^{n_{\text{issues}}} 
    \left( \sum_{c=1}^{n_{\text{choices}}} p_{c,i,l} \ln p_{c,i,l} \right)
\end{equation}   

where $p_{c,i,l}$ denotes the current frequency that an agent holding label $l$ expresses choice $c$ regarding issue $i$ .
To measure the likeness between choices of agents from two different labels on a specific issue, we examine the distribution of agent support for each possible choice within both groups. Then, we calculate cosine similarity between resultant vectors representing frequencies of choices. A similarity score of 1 denotes identical vectors (global alignment), while 0 means that they are orthogonal (global misalignment). Subsequently, we compute the average similarity across issues and all possible pairwise comparisons. The mathematical formula to calculate the Cosine similarity index is presented in equation 2:

\begin{equation}
    \textit{Cosine index} = \frac{1}{n_{\text{issues}}} 
    \sum\limits_{i=1}^{n_{\text{issues}}} 
    \frac{1}{2 n_{\text{labels}} (n_{\text{labels}} -1)} 
    \sum\limits_{l_1=1}^{n_{\text{labels}}} 
    \sum\limits_{l_2=l_1+1}^{n_{\text{labels}}} 
    \left( \frac{\sum\limits_{c=1}^{n_{\text{choices}}} P_{c,i,l_1} \cdot P_{c,i,l_2}}
    {\sum\limits_{c=1}^{n_{\text{choices}}} \sqrt{P_{c,i,l_1}^2} \sum\limits_{c=1}^{n_{\text{choices}}} \sqrt{P_{c,i,l_2}^2}} \right)
\end{equation}  

where $p_{c,i,l}$ is defined as in equation 1.

\section{Results}

We conducted multiple simulations across a broad range of parameter settings (Table 1) to examine the dynamics produced by our model of political opinion dynamics, which is grounded in hierarchical coarse-graining and community-bounded social influence. The results indicate that the model robustly generates large-scale patterns of information integration and alignment within and across labels. Notably, the parameter space segregates into three distinct regimes of information integration, predominantly modulated by the degree of higher-level connectivity. These regimes are distinguished by differences in the evolution of alignment and misalignment within labels, as measured by the Shannon-Wiener diversity index, and at the global level, as measured by cosine similarity. As discussed in Section 3.3, we focus on a narrower parameter range that remains qualitatively consistent with these regimes to illustrate our findings.

\subsection{Regimes of Information Integration}

Our results reveal three distinct regimes of information integration, determined by the probability of label switching (prob-dropping-label, PDL) as well as ‘ignoring’ off. Each regime exhibits a unique pattern of choice integration both within and across labels, influencing the system's progression toward local and global consensus. Below, we provide a detailed description of each regime. To enhance clarity and coherence, we have restructured the results and discussion sections. Consequently, Figures 5 and 6 contain examples that do not directly correspond to the regimes they describe. Each section will explicitly clarify these instances.

\subsubsection{Independent Information Integration Regime}

When \textit{prob-dropping-label}, (PDL) equals 0, no individual switches across labels, leading to a rapid alignment of choices within the label but no integration across labels. In this case the diversity of choices within labels decreases (SW index quickly drops close to zero), indicating convergence of choices within labels (Fig. 5b). Further interactions do not lead to any choice changes after timestep \textasciitilde500, indicating equilibrium. After timestep 0, the diversity across labels is high (Cosine of \textasciitilde1), increases until timestep \textasciitilde400 (Cosine index drops) and remains at a stable level (Fig. 5a). As a result, the system reaches an early steady state, where labels maintain distinct and isolated choices, forming persistent informational silos that prevent global convergence. We refer to this as the independent information integration regime, where no information flows between labels, leading to global misalignment but label alignment (Fig. 5). While the figure is intended to illustrate the independent regime, it includes some settings where ignoring is off, that do not align with this regime. 

\begin{figure}[H]
    \centering
    \begin{subfigure}{0.49\textwidth}
        \centering
        \includegraphics[width=\textwidth]{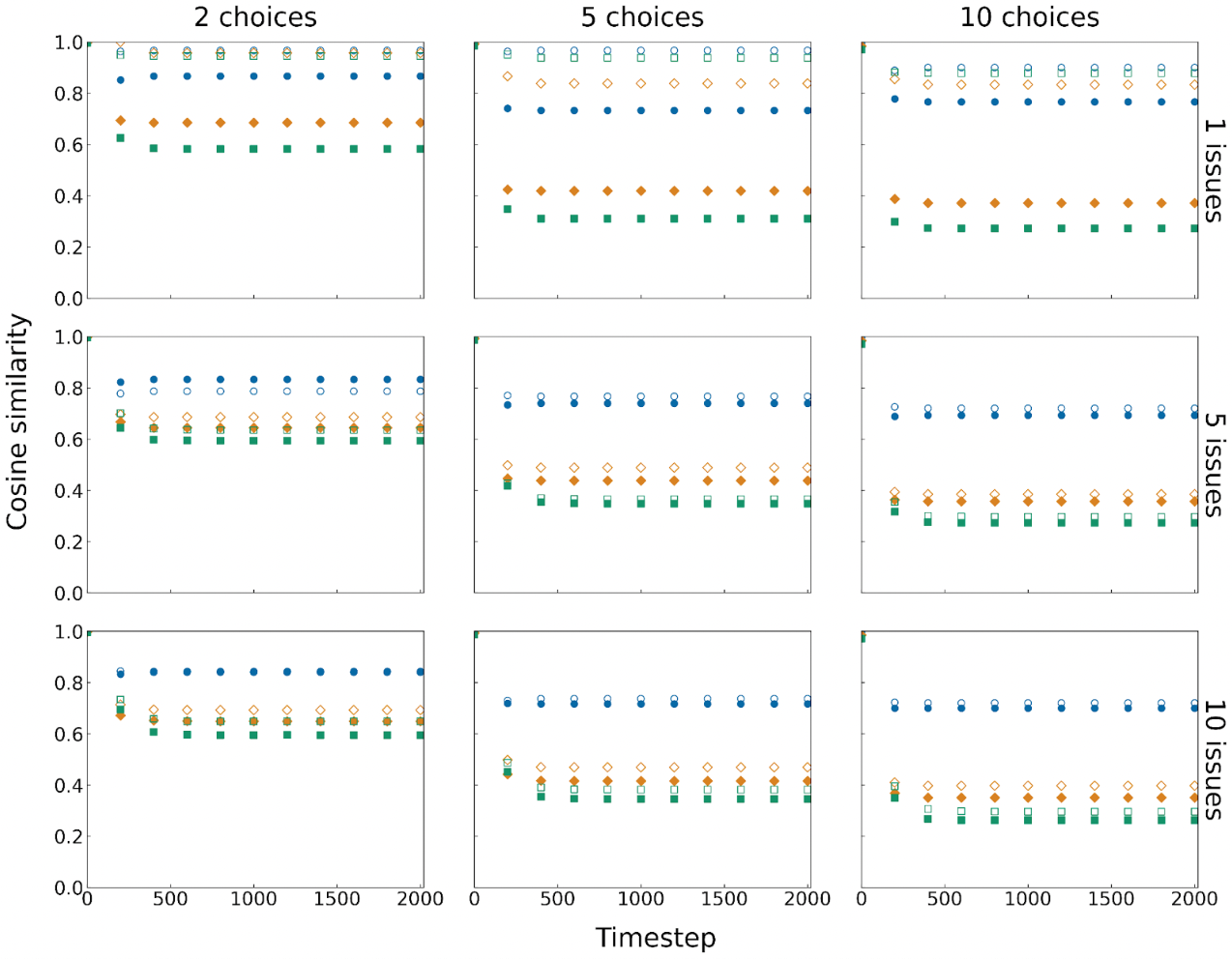}
        \caption{Cosine similarity}
        \label{fig:cosine5}
    \end{subfigure}
    \hfill
    \begin{subfigure}{0.49\textwidth}
        \centering
        \includegraphics[width=\textwidth]{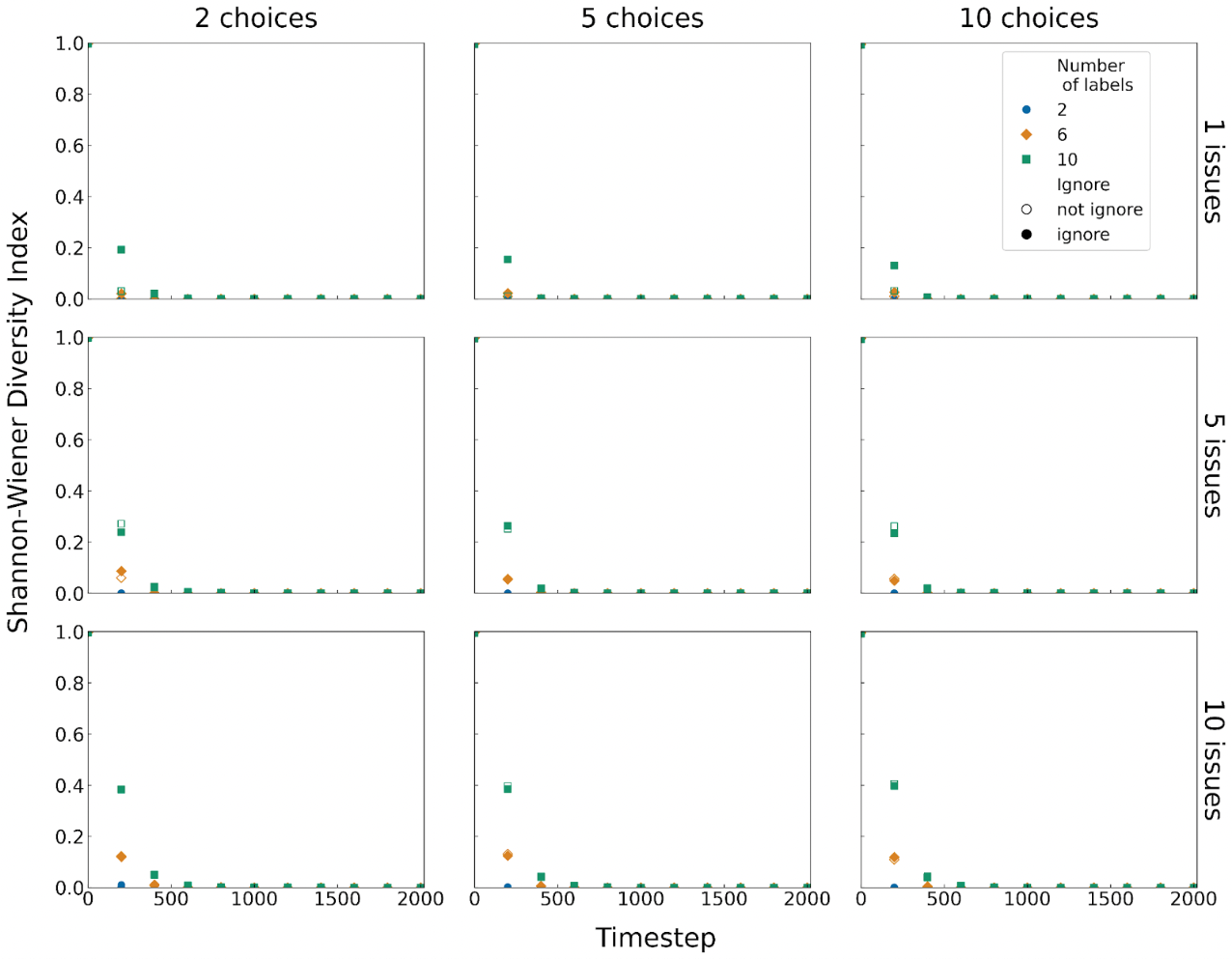}
        \caption{Shannon-Wiener diversity}
        \label{fig:shannon5}
    \end{subfigure}
    \caption{Independent information integration regime. Results shown are means of 10 simulations per parameter set. PDL = 0; Strength of influence = 20; multi-issue discourse = 20; and ignore is ON or OFF (filled and hollow points). 2, 6, or 10 labels (color and shape-coded points); 1, 5, and 10 issues (rows); and 2, 5 or 10 choices (columns)}
    \label{fig:combined5}
\end{figure}

\subsubsection{Parallel Information Integration Regime}

When PDL is sufficiently high (>= \textasciitilde0.001, depending on the exact configuration of issues, labels, choices, SoI and MID), agents switch labels more frequently. In this regime, the Cosine index starts completely aligned and quickly drops, indicating divergence, from which point on the Cosine index increases until global alignment is reached (timestep \textasciitilde500). Within labels, we can observe a convergence of agents’ preferred choices, however at a slower rate than global alignment, reaching local alignment for all parameter settings between timestep \textasciitilde1000 to \textasciitilde2000 (Fig. 6b). The fact that global alignment precedes label alignment suggests that agents are integrating information globally at such a rate that an initial divergence between labels is eliminated before alignment within labels is reached. Because global and local alignment are reached in the same phase, we refer to this as the parallel information integration regime. The high variance in label alignment trajectories (Fig. 6b), illustrates that the PDL of 0.01 in combination with various parameters produces dynamics that border on the iterative information integration regime (see below). In the case of two labels and ignoring on for all parameter settings, as well as ignoring off for 5 and 10 issues (Fig. 6b), we can see transient increases in diversity after initial local alignment, clearly depicting an iterative regime. 

\begin{figure}[ht]
    \centering
    \begin{subfigure}{0.49\textwidth}
        \centering
        \includegraphics[width=\textwidth]{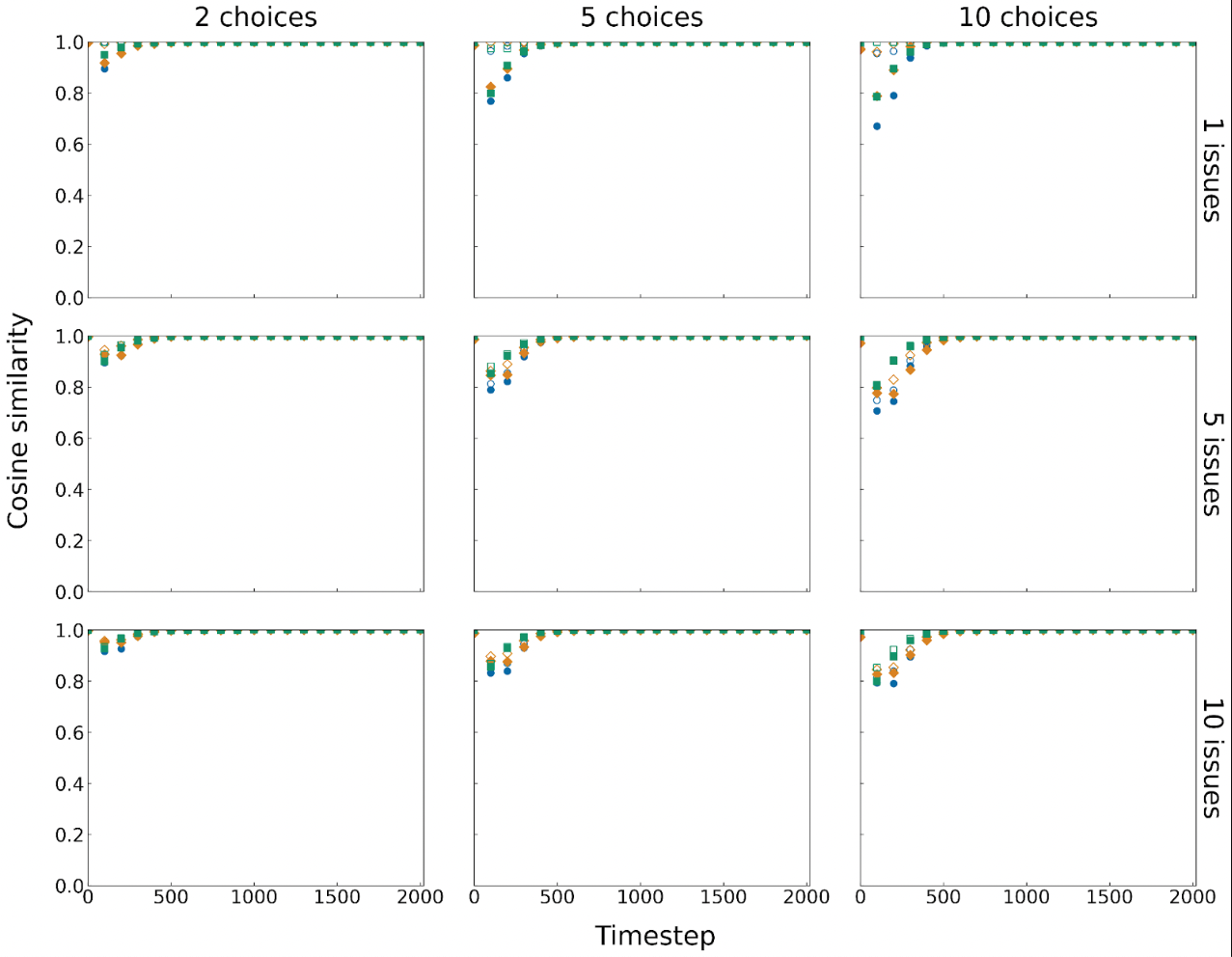}
        \caption{Cosine similarity}
        \label{fig:cosine6}
    \end{subfigure}
    \hfill
    \begin{subfigure}{0.49\textwidth}
        \centering
        \includegraphics[width=\textwidth]{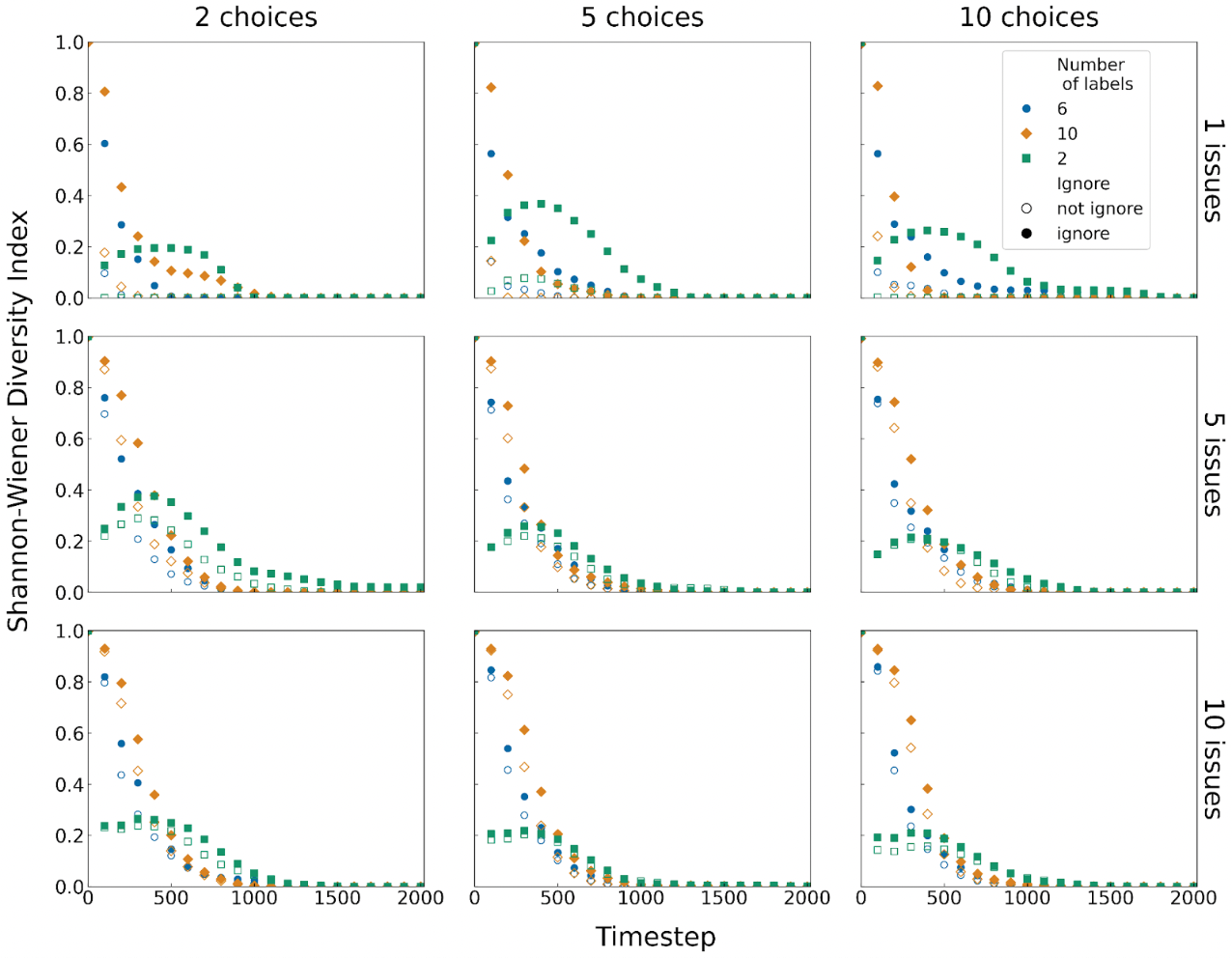}
        \caption{Shannon-Wiener diversity}
        \label{fig:shannon6}
    \end{subfigure}
    \caption{Parallel information integration regime. Results shown are means of 10 simulations per parameter set. PDL =0.01; Strength of influence = 20; multi-issue discourse = 20; and ignore is ON or OFF (filled and hollow points). 2, 6, or 10 labels (color and shape-coded points); 1, 5, and 10 issues; and 2 or 10 choices (columns)}
    \label{fig:combined6}
\end{figure}

\subsubsection{Iterative Information Integration Regime}

In the independent and parallel regimes, a final steady state is reached relatively quickly, within 500-2000 timesteps (Fig. 5-6). These regimes show rapid increases in alignment, either within labels only (independent regime) or at both global and local levels (parallel regime). A different regime emerges under intermediate levels of PDL (0 < PDL < \textasciitilde0.005, depending on the exact configuration of issues, labels, choices, SoI and MID), where rapid and temporary local alignment is followed by an increase in divergence of preferred choices within labels until a ‘transient diversity’, i.e. temporary misalignment, within labels is reached. This is clearly illustrated in Fig. 7b, where the SW index drops rapidly, indicating aligning preferred choices within labels. A high Cosine index (Fig. 7a) shows that this state of local alignment coincides with global misalignment. This is followed by a temporary increase of divergence at label level. The Cosine index seems to show that global alignment is reached while labels are once more internally misaligned (Fig. 7b). Equilibrium (global consensus) which encompasses both global and local alignment for the parallel and the iterative regime, is reached only after \textasciitilde10,000 timesteps, making it considerably longer than the parallel regime. 

\begin{figure}[ht]
    \centering
    \begin{subfigure}{0.49\textwidth}
        \centering
        \includegraphics[width=\textwidth]{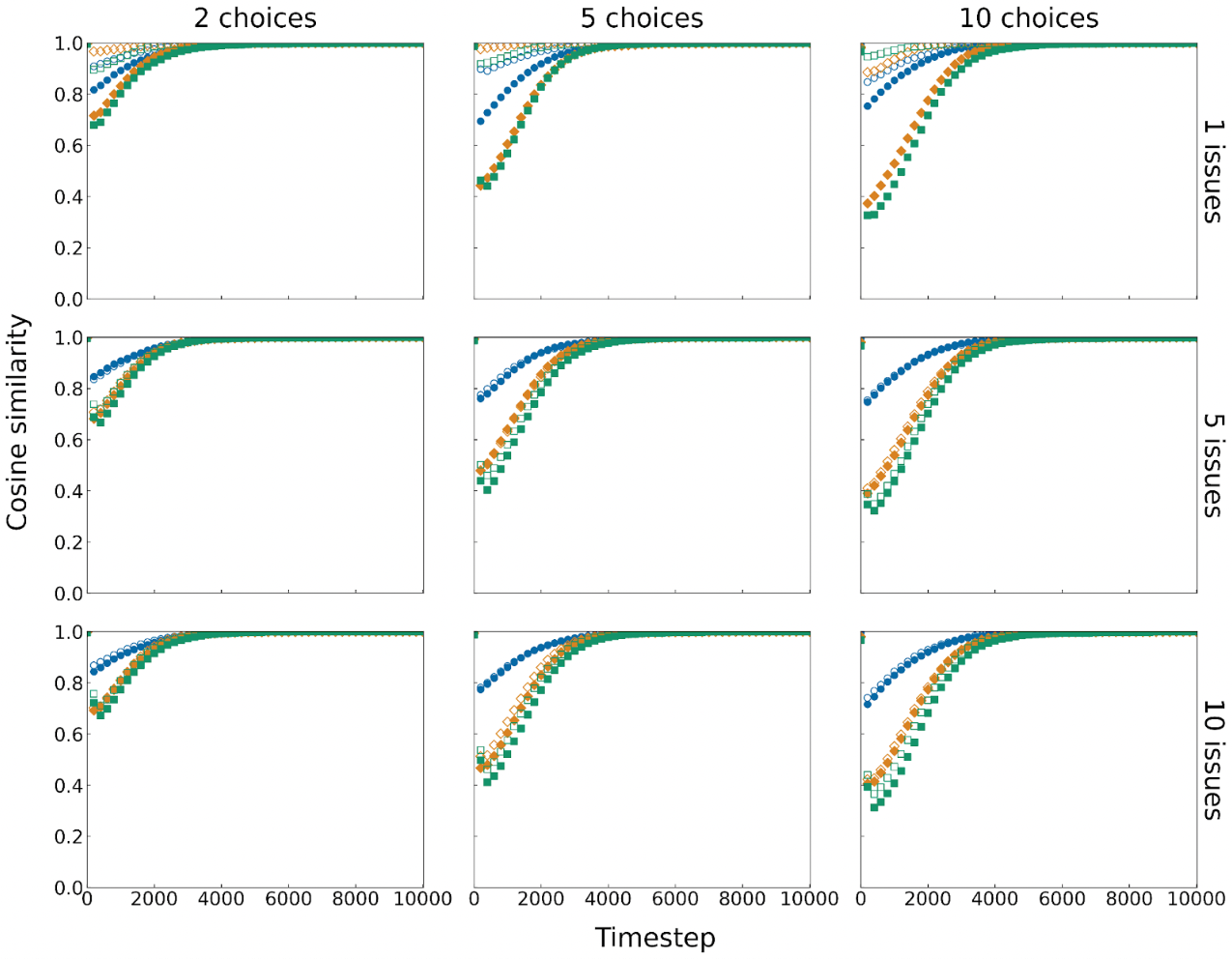}
        \caption{Cosine similarity}
        \label{fig:cosine7}
    \end{subfigure}
    \hfill
    \begin{subfigure}{0.49\textwidth}
        \centering
        \includegraphics[width=\textwidth]{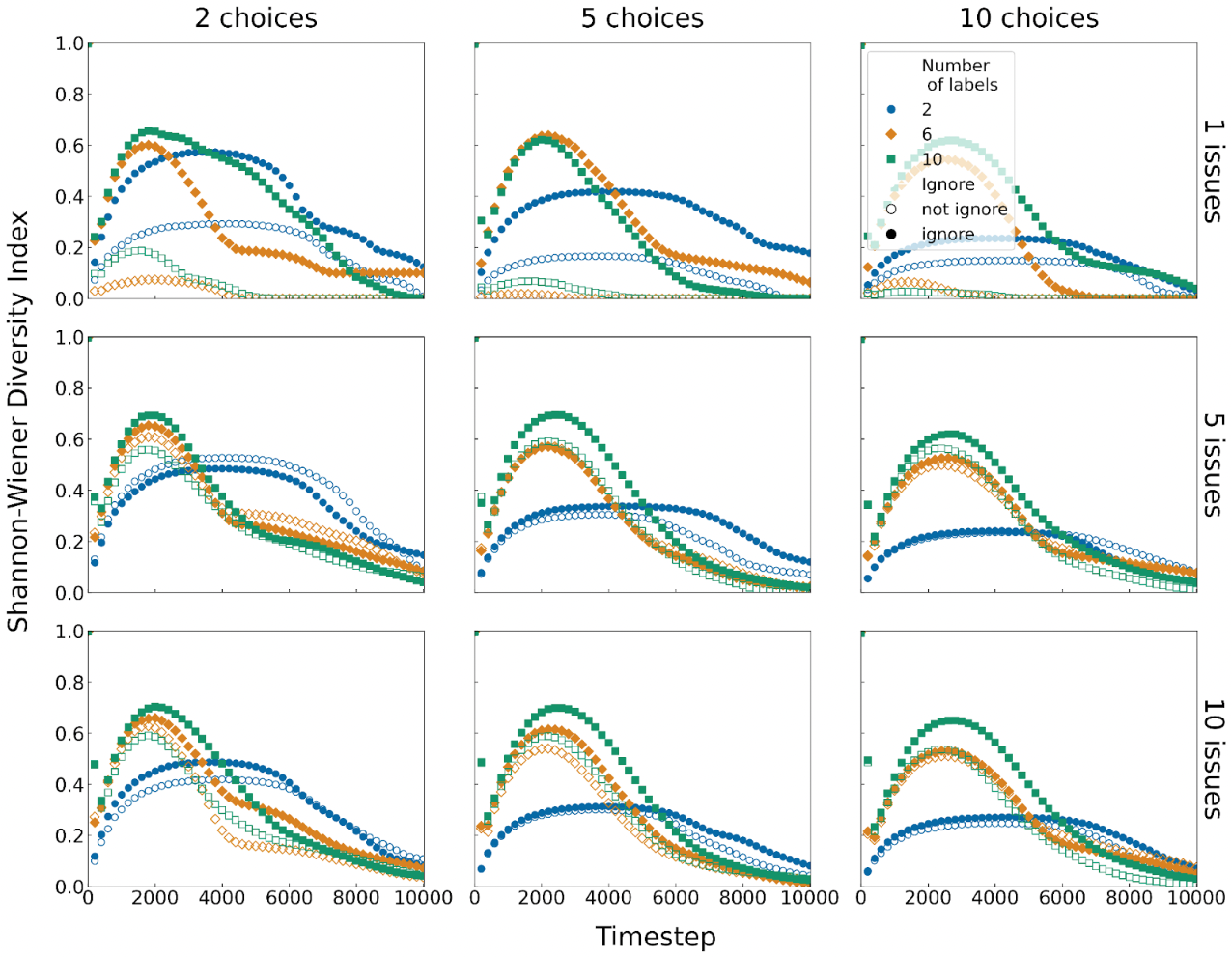}
        \caption{Shannon-Wiener diversity}
        \label{fig:shannon7}
    \end{subfigure}
    \caption{Iterative information integration regime. Results shown are means of 10 simulations per parameter set. PDL = 0.001; Strength of influence = 20; multi-issue discourse = 20; and ignore is ON or OFF (filled and hollow points). 2, 6, or 10 labels (color and shape-coded points); 1, 5, and 10 issues; and 2 or 10 choices (columns)}
    \label{fig:combined7}
\end{figure}

The SW index shows that the iterative regime gives rise to transient diversity at the local level and has three phases: local alignment, local divergence/global alignment and local alignment. We refer to this as the iterative information integration regime because the initial alignment within labels gives rise to a global misalignment, starting a separate information integration between higher-level components (the labels) until global alignment is achieved. In the next section, we focus on the phases of a specific instance of the iterative regime and analyze the underlying causation (Fig. 8,9,10). However, note that the patterns described below are qualitatively consistent across all simulations with an iterative information integration regime.

\subsection{Analysis of Hierarchical Causation in the Iterative Regime}

In order to understand what ultimately leads to the information integration phases and regimes we have observed, we employ the previously defined indicators for upward and downward causation to a simulation run of the iterative regime. In addition, we rely on the activation barrier concept, as defined in the background, to support our analysis.
To identify the different phases and track the alignment or misalignment of the distributions, we plotted the SW index and Cosine index of an iterative regime simulation run.

\begin{figure} 
    \centering
    \includegraphics[width=1\textwidth]{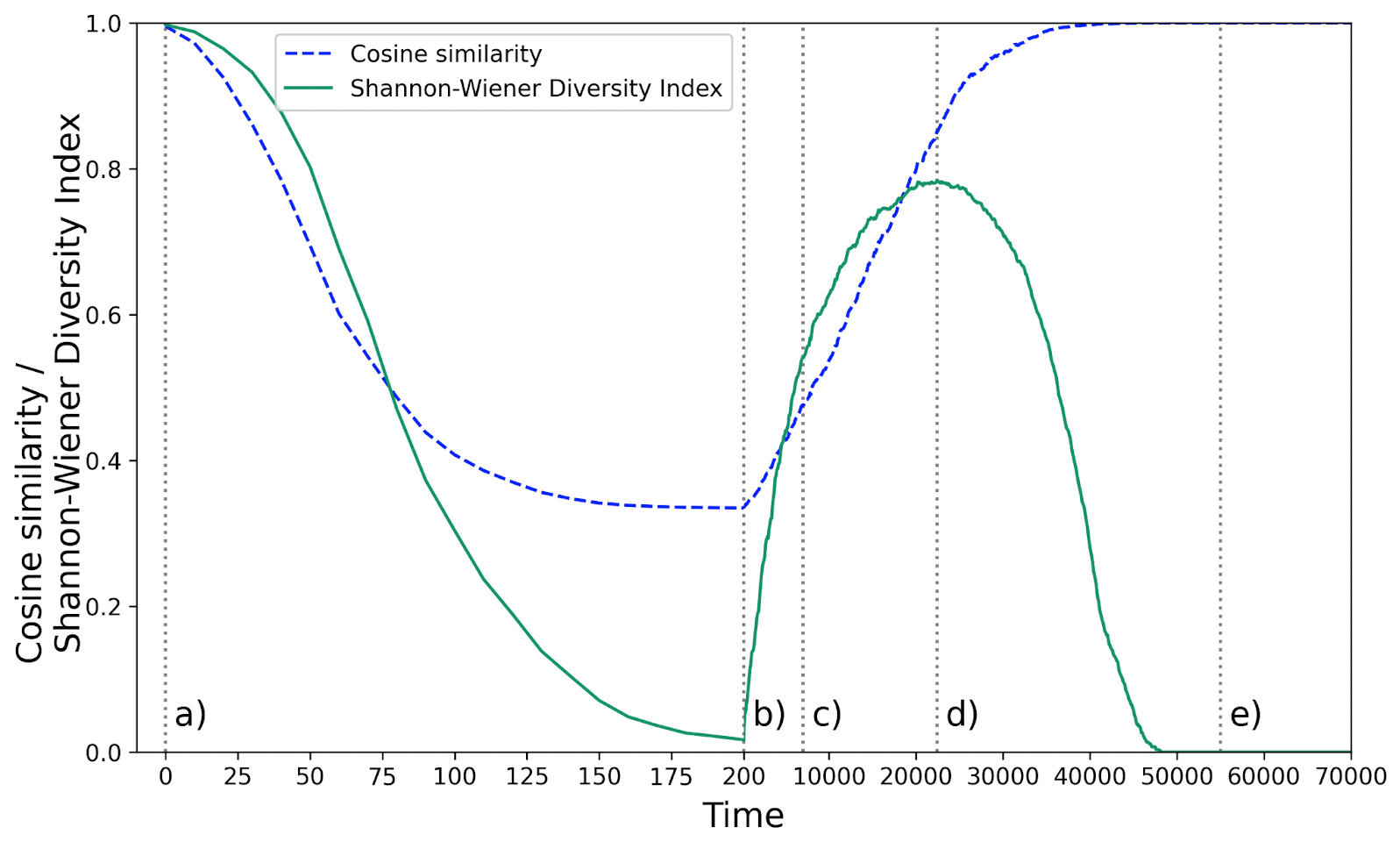}
    \caption{Iterative information integration regime. 
The time-scale (x-axis) is adjusted to show the decrease of relative entropy in Phase I in more detail. PDL = 0.0001; Strength of influence = 20; multi-issue discourse = 20; and ignore is ON; 8 labels; 1 issue; and 5 choices. Different timepoints are depicted by the dashed vertical lines. Point a = 0, b= 200, c = \textasciitilde7,000; d= \textasciitilde22,400; and e = \textasciitilde55,000 timesteps}
 \end{figure}

We further mapped the distributions of preferred choice for each level (agent, label, global) in all the phases of the iterative regime to identify changes in activation barriers, downward causation and alignment. Fig. 9, outlines the distributions at the agent and label level for label 1. Columns 1-5 show the choices’ strength distribution for agents preferring choice 0-4 (columns) at different time points (rows). The last column to the right shows the label’s distribution of agents with a given preferred choice at different timesteps.

\begin{figure} 
    \centering
    \includegraphics[width=1\textwidth]{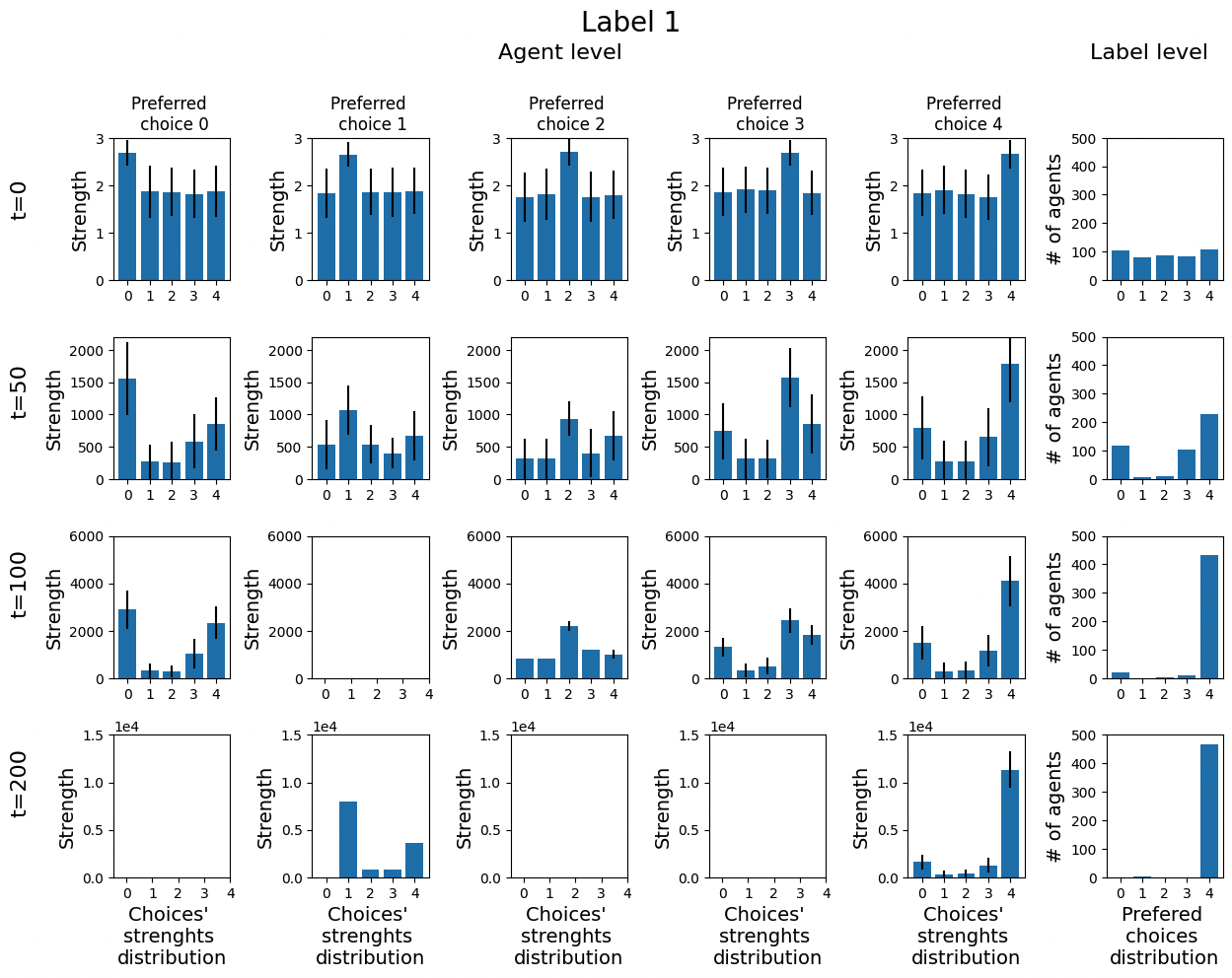}
    \caption{Distributions of choices’ strengths at the agent level and label choice distributions of Label 1 at different time points (rows) during Phase I. In the agent-level panel, columns represent agents preferring a given choice, bar plots show the average strength of choices, and error bars indicate standard deviation. Blank plots indicate that no agents preferred the given choice. The label-level panel (last column on the right) presents bar plots depicting the number of agents preferring each choice at different time points. PDL = 0.0001; Strength of influence = 20; multi-issue discourse = 20; and ignore is ON; 8 labels; 1 issue; and 5 choices}
 \end{figure}
 
\subsubsection{Phase I: Local Alignment}

At timestep 0, the activation barrier to preferred choice change is low (Fig. 9). The SW index is highest (Fig. 8, point a), due to the normal distribution at initialization. After interactions begin, at timestep 50, we can see that all choices are still preferred by at least one agent in the label (last column, row 2 in Fig. 9). The relative entropy among choices is decreased and the activation barrier goes up from 2 at timestep 0, to \textasciitilde500 at timestep 50 and \textasciitilde10000 at timestep 200 (see y-axis values at different time steps in Fig. 9). The label preferred choice distribution increasingly converges: By timestep 50, choice 1 and 2 are almost eliminated; by timestep 100, choice 1 is eliminated and agent’s preferring choices 0, 2, or 3 are on the verge of switching preferred choice; by timestep 200, it is clear that the agents will converge on choice 4. This corresponds to point b in Fig. 8. We can see that the preferred choice distribution stabilizes around timestep 50 (consolidating around preferred choice 4) which triggers downward causation leading all agents to converge on choice 4. The rate at which agents change their preferred choice shows a sharp decline after \textasciitilde10 timesteps (Fig. 46 in the supporting information \cite{Kaminska2025}, the local level becomes time-scale separated. Hence, the preferred choice at timestep 50 predicts the future state of the agents at timestep 200, and the mutual information between the label and the agents increases. All agents tune to the same social environment given by the model design that restricts information integration to the local community for all members of a label. All of Flack’s \cite{Flack2017a} outlined criteria for downward causation are thus observable in phase I.

We argue that the time-scale separation and stability of the local distribution results in local downward causation because it changes the inputs to agents. It increases the precision of the input term in depicting the local distribution across agents, producing coherent output in the direction of the locally preferred choice. We believe this is a consequence of the increased amount of samples needed until agents can change their preferred choice due to the activation barrier (Fig. 9). A perfectly accurate coarse-grained estimation of the higher-level distribution can only result in that local choice. If an agent were to sample everyone else in their community before their preferred choice changes, and in the amount of time they are doing so the distribution remains roughly the same, the input term would have to be highly accurate in conveying the real local distribution, which is what agents are estimating \cite{Flack2017a}. This would lead the agent to change its state to the preferred choice of the label. The more agents become accurate in their estimation, the smaller the pool of possible outcomes, making agents in the label converge on the locally preferred choice. Therefore, an activation barrier, increasing evenly in agents of a local community and globally, can be thought of as increasing the accuracy of the input term that serves as the measurement of the local distribution for each agent. In a nutshell, the local distribution - due to a high level of accuracy of the distribution of inputs until opinion change in measuring the local distribution - causes (downward causation) the output (agent upward causation) and in turn the agents’ preferred choice aligns with that of the local distribution. As can be seen in Fig. 10, ts = 200, the iterative regime gives rise to global misalignment after phase I. At higher levels of PDL, the parallel information integration of the global distribution can lead to global alignment before local alignment (parallel regime). Since all communities tune to the same global distribution, the stabilization of the global level with local alignment and remaining global misalignment activates downward causation.

\begin{figure} 
    \centering
    \includegraphics[width=0.95\textwidth]{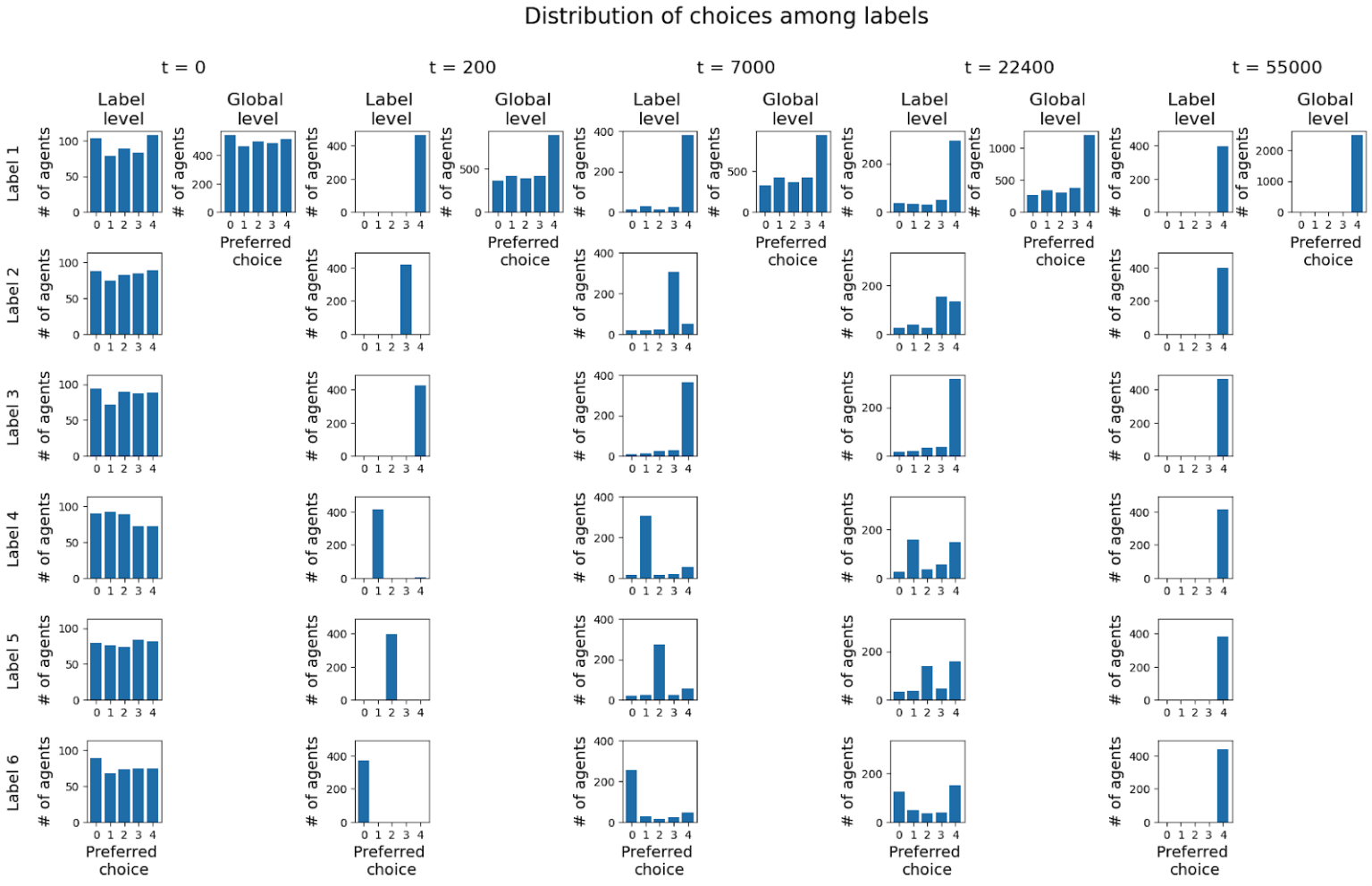}
    \caption{Preferred Choice distributions at the label and global level at different time points during stage I-III. Rows represent different labels and histograms represent the number of agents preferring a given choice at the label or global level. PDL = 0.0001; Strength of influence = 20; multi-issue discourse = 20; and ignore is ON; 8 labels; 1 issue; and 5 choices}
 \end{figure}

\subsubsection{Phase II: Global Alignment}

Fig. 10 outlines the preferred choice distribution at the label and global level at different time points during the simulation. At the beginning of phase II (point b in Fig. 8 and t=200 in Fig. 10), the labels are internally aligned, but there is global misalignment, even though there appears to be a preference towards choice 4. At this point, two labels (1 and 2) are aligned with the global choice but all others prefer a different choice. These four divergent labels (2, 4-6) switch or start switching to choice 4 around timestep 22,400, point d in Fig. 8. Like in phase I, the switching of preferred choices is asynchronous and divergence at the global level decreases one label at a time. Note that the preferred choice at the global level remains unchanged from timestep 200 on until the end of phase III, and in phase II the global distribution undergoes only marginal changes, whereas labels diverge in their preferred choice distribution till timestep 22,400 (point d in Fig. 8, end of phase II). We can identify global downward causation because the global distribution, that becomes stable at point b (Fig. 8), compels labels to align their preferred choices with the global preference (choice 4) at point d (Fig. 8), resulting in marked global alignment and a significant increase in mutual information. 

During this phase, agents are sampling and reinforcing the preferred choice of their label (local downward causation), contributing to the increase in the agent’s activation barrier, while also sampling the choices of migrating agents (global downward causation), likely diverging from the label’s preferred choice. We argue it is this moving of agents with high activation barriers (stubborn and high in certainty) across labels that increases the diversity of preferred choices within labels and ultimately aligns labels with the global distribution. The stability of the global distribution stems from the fact that it is constituted of labels, and the only source of change in the global distribution comes from labels changing their preferred choice (label upward causation). In phase II, labels themselves have a high activation barrier (stubborn agents and highly aligned distribution) but compared to the global distribution they are more reactive to the information input. Diverging agents from other labels keep coming in, and eventually, the divergent information crosses, first, the agents’ activation barrier and then, the label’s activation barrier to preferred choice change. Therefore, the label tunes to the global opinion distribution and not the other way around. However, slowly, since the ratio between label and non-label agents at every timestep is tipped in favor of the label. 
The outcome of phase II is global alignment of preferred choice and label misalignment. Global and local downward causation are aligned, having the same preferred choice, and as a result the local distribution is not subject to further divergence.

\subsubsection{Phase III: Local Alignment}

In our example, this phase occurs between timestep 22,400 (point d, Fig. 8) and 55,000 (point e, Fig. 8). The alignment of the local and global social environment (due to global alignment from the previous phase) signifies that the agents, again, coherently tune to the same local distribution and stable preferred choice, mirroring the conditions of local alignment during phase I. In phase III, agents increasingly converge in their estimates of the label’s probability distribution until local alignment, signifying consensus of all agents in all labels with the coarse-grained preferred choice 4 (Fig. 10, t = 55,000). At this point, the system is in equilibrium: every encountered agent agrees on the preferred choice (local and global alignment) and the continued interactions cease to cause preferred choice changes, only serving to increase agents’ activation barriers.

\subsubsection{Hierarchical Coarse-Graining Mechanism with Regimes}

The causal analysis, combined with our previous observations, can be abstracted into an algorithmic mechanism illustrated in Fig. 11. In Phase I, the system transitions from initial local and global misalignment to either complete alignment (the parallel regime), solely local alignment (the independent regime), or, under higher-level connectivity, global misalignment initiates a second phase of the iterative regime. In Phase II, the diversity of preferred choices across labels converges toward a single global preference (Outcome II, Fig. 11), which in turn disrupts local alignment (Outcome I, Fig. 11). Finally, Phase III starts with local misalignment and ends with local alignment (Outcome I, Fig. 11), which in combination with the global alignment yields global consensus, thereby fulfilling the end condition of the iterative regime.

\begin{figure} 
    \centering
    \includegraphics[width=0.80\textwidth]{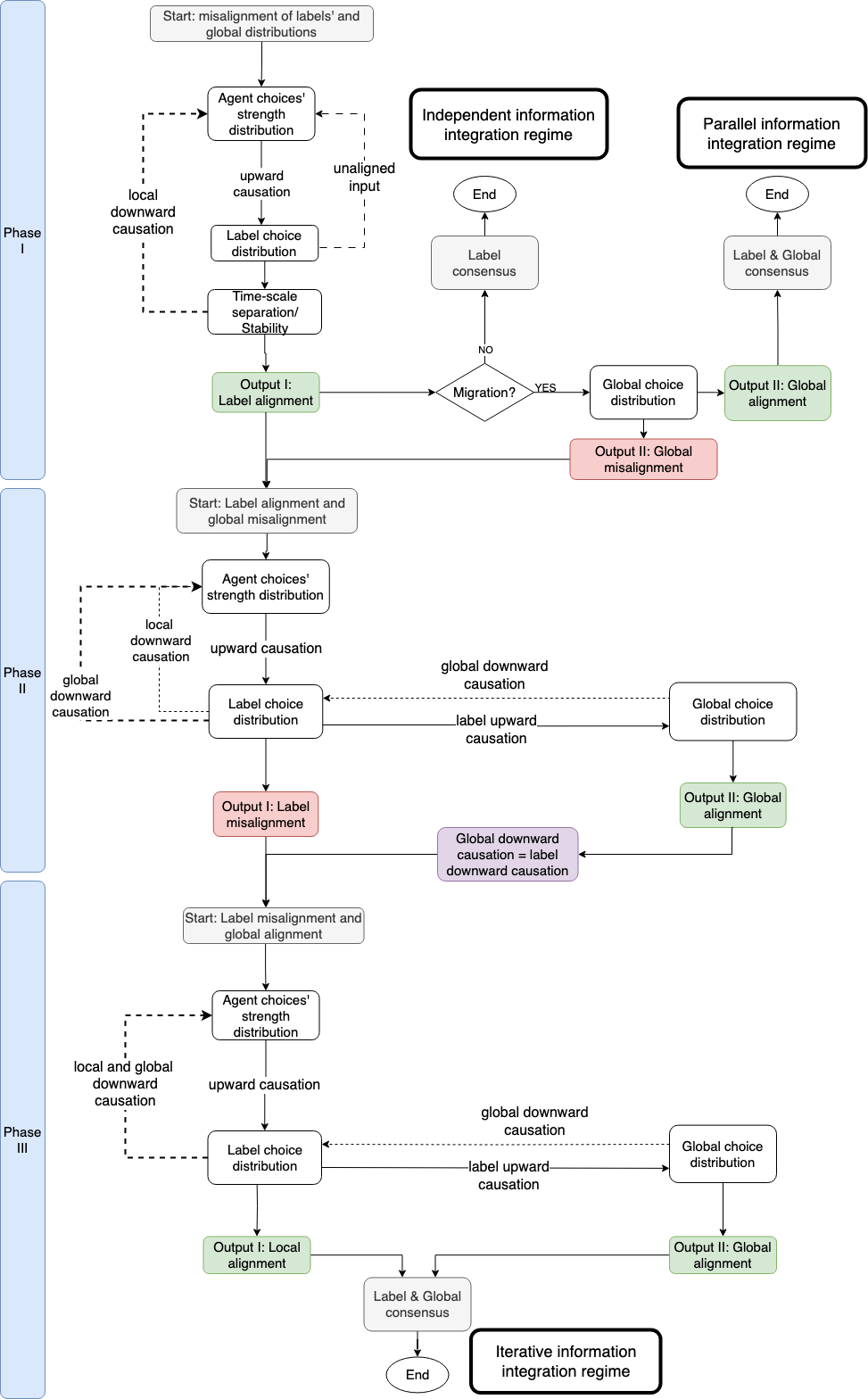}
    \caption{Hierarchical coarse-graining mechanism with regimes. Phases start with the output of the previous one: local and/or global misalignment, activated downward causation; green: alignment, red: misalignment; process ends when all misalignment in the social environment is resolved}
 \end{figure}

\section{Discussion}

Our paper addresses the gap in our understanding of what causal mechanism of information integration leads to the emergence of consensus in political opinion dynamics and polarization, a dysfunction of the information integration mechanism. In combining an understanding of political opinion dynamics with a hierarchical coarse-graining mechanism, we were able to trace the emergence of downward causation, higher levels of organization and consensus in a multilevel mechanism. A realistic account of such a causal mechanism is necessary to inform further research and policy interventions that can successfully curb polarization. Hierarchical causation helps explain how individual opinions aggregate into collective beliefs (upward causation) and how those collective beliefs, once stabilized, shape future individual opinions and alignment (downward causation). In this section, we examine the broader implications of our model results in the context of polarization.

Our results show that social influence based on shared social identity alone confines information integration to the local community (independent regime). The implementation of noise, simulating higher-level information integration between local communities, leads to consensus at any value of label switching probability. A low value of probability leads to an iterative regime that is distinguished by a transient diversity before global consensus. The iterative regime has the additional phases of global alignment (phase II) and local alignment (phase III) and is substantially longer than the other regimes. This might be universally beneficial for democratic functioning since overall more information contributes to the final slowly constructed variable \cite{Smaldino2022, Flack2017a} and point to a general benefit of a low degree of higher-level connectivity. A higher value of switching probability leads to a parallel information integration regime with a rapid, and nearly simultaneous local and global alignment. The parallel regime exhibits fast variable construction, which seems to respond to ‘herding’ \cite{Smaldino2017}. Public discourse is commonly understood as the discussion of diverse points of view on a given issue at the global societal level, with opinions characterized by some degree of certainty about one’s perspective before agreement. Therefore, the iterative information integration regime which includes a phase II of high personal certainty and diversity of points of view, corresponds most closely to public discourse in democracies.

Downward causation in our model arises dynamically from emergent time-scale separation. It first comes about in phase I between micro- and meso level due to the increasing barrier to opinion change. At the end of phase I, when labels are fully aligned and the activation barrier to preferred choice change at the local level is high, the global distribution becomes consolidated. Time-scale separation stabilizes the distributions, providing coherent and accurate input to components at the lower scale. Agents start to converge when their estimates of the local choice distribution start to align, i.e. tuning becomes increasingly coherent across agents (Flack, 2017), and the number of agents with the most popular choice of that label starts to increase. In a nutshell, the causation needs to become “circular” \cite{Fuchs2020}, where the higher-level distribution that exerts downward causation becomes itself affected by its effect on agents changing their choice, which in turn converges the higher-level distribution itself. Fittingly, Fuchs calls circular causation “downward/upward causation or global-to-local/local-to-global causality” \cite{Fuchs2020}. By thinking about what real world phenomena could intervene in the emergence of circular causation, we can hypothesize what might cause democratic societies to struggle with coming to consensus. Since information connectivity in democratic societies beyond local communities is high, the independent regime can be ruled out. An inability to come to global consensus thus indicates a failure to complete coarse-graining by not meeting the emergent conditions for local or global downward causation. The discovered mechanism of our model can give insight into how ideological polarization, which is a disagreement on policy positions, and affective polarization, ingroup favoritism and outgroup disdain, may arise in democratic societies.

According to Hobolt et al. (\cite{Hobolt2024}, polarized societies exhibit political conflict that “solidifies and political identities crystallize into polarized groups” \cite{Hobolt2024}. Further, affectively polarized societies show high homogeneity within the polarizing communities and sustained disagreement and disdain between them \cite{Bednar2021, tornberg2022}. This is akin to locally aligned groups interacting but maintaining their preferred choice nonetheless, sustaining a situation of global misalignment. A common explanation ascribed to continued disagreement and hostility between opinion groups is the idea of echo chambers, or social environments with homogenous opinion distributions, that confirm each other’s point of view \cite{Barbera2020, Hobolt2024}. In the iterative regime, after phase I, most members of labels have converged on a single choice which they keep reinforcing in interactions. Agents are stubborn and high in certainty, so the activation barrier is high. The label at this point can be compared to an echo chamber, in that choices are homogeneous and interactions continuously amplify the shared choice. Interestingly, even while agents within labels keep confirming their choice, the inter-community information integration from migrating agents inevitably leads to the divergence of preferred choices within the label and the alignment of labels over time, conflicting with the ascribed role of an echo chamber in entrenching ingroup views. In fact, the temporary creation of a locally aligned social environment is positive in our model, in that it is the precondition for global time-scale separation and thus global downward causation. What allows for the local alignment pressure exerted by local downward causation to be overcome, is the \textit{global} downward causation that emerges when a stable and time-scale separated global distribution transmits information to all local communities equally. It is important to note that by design local communities in our model tune to the same global distribution that real-life political opinion dynamics might not.

Applied to the echo chamber phenomenon, this would signify that any exposure and integration of information that comes from a more stable (i.e. a higher-level distribution with time-scale separation) source will lead to information integration and alignment across communities, if the information source is the same for all local communities comprised in the higher-level distribution. This suggests that what matters is not only, as previously thought, the similarity of the input of community members \cite{Hobolt2024} but also the input the community receives from the wider social network it is embedded in. More specifically, the input distribution all local communities are tuning to should possess the same majority or preferred choice and as in our model, time-scale separated from the level below, such that global downward causation can emerge and produce alignment across communities around that global preferred choice. What might affect time-scale separation and coherence of preferred choice of the inputs into local communities in democratic societies? 

In our model, higher-level connectivity, i.e. the flow of information between communities, is implemented as a type of noise on the social influence heuristic that affects communities equally. The global distribution that consists of the communities and their preferred choices provides the same reference frame for each community. A source of incoherence of inputs might arise from the rejection of a particular community’s opinion by another or when a local community rejects a certain opinion or choice. However, this only prevents global alignment when the ignored portion of the global distribution shifts the global preferred choice from the perspective of the ignoring local community, leading this particular community to align with their perceived global majority choice distinct from the other communities. This is possible when the communities are not equal in size and the ignored one tips the majority. When the criterion for ignoring is the choice itself, then all communities favoring that choice are ignored. This necessarily means that the ignored opinion is the majority opinion to produce a shift and therefore prevent alignment of that community with the actual global preferred choice. Other pathways for communities to perceive a different global preferred choice would be a categorical rejection of the majority or through the mediating information architecture. 

In 21st century democratic societies, people do not only talk to their ingroup to reduce their uncertainty about an issue. They also consult sources of information that are broadcast from the top-down, such as TV shows, the radio, the newspaper or the internet that transmit the opinions of experts, politicians and the general public. The trust an individual has in the information source allows information from beyond the local community to influence them. When these globally accessible sources of information are trusted by all local communities, we can postulate that they are embedded in a fairly coherent information environment. Conversely, a source of incoherence would arise from community-based differences in what constitutes a trustworthy source of information, which agrees with findings that trust in media is highly polarized in the U.S. \cite{tornberg2022}. Further, research into the connection between misinformation and affective polarization provides evidence that people who are affectively polarized trust information that is aligned with their ingroup and distrust information that is aligned with the outgroup \cite{Jenke2023}.

The finding that incoherent inputs across agents and communities hinder global consensus in political opinion dynamics aligns with Durkheim’s observation that consensus on opinions and values requires social cohesion, which emerges from individuals’ exposure to highly similar social influences \cite{Durkheim1982}. Political theory on social cohesion states that social conflict is an inherent feature of a healthy society as long as lines of division are overlapping, i.e. social groups might be allied or opposed, depending on the issue, creating a plural and cross-cutting tapestry of political conflict \cite{Coser1956, tornberg2022}. The theory of ‘partisan sorting’ argues that affective polarization is related to a process of aligning such cross-cutting social, cultural and ideological divides into two opposed mega-identities along partisan lines \cite{tornberg2022}. According to Törnberg (\cite{tornberg2022}) sorting is driven by non-local interactions in digital media that expose individuals to a wide range of political opinions and attached social identities that they agree and disagree with. He argues that this exposure amplifies perceived differences with outgroups and similarities within ingroups, restructuring how individuals categorize themselves and others within the social landscape. This realignment of social influence and cross-cutting divisions converges into a singular salient identity, the political dimension, leading to a breakdown of cross-cutting global social cohesion, isolating information flows to the political identity across an expanding range of issues \cite{tornberg2022}.

A second implication of the discovered coarse-graining mechanism for affective polarization emerges in phase I, when local downward causation must develop to achieve initial local alignment. When agents have non-local inputs in this first phase, the conditions for local downward causation may not emerge. Depending on the amount of online versus real-life interaction and the degree of individualization of information input into the community, the requirement of coherent tuning to the same aggregate distribution and preferred choice might be violated. An incoherent input across the local community might thus block downward causation and local alignment from arising, resulting in individuals aligned with non-local communities. This relates directly to Törnbergs (\cite{tornberg2022}) finding that non-local interactions impact feelings of similarity and difference, potentially leading to new overarching social identities as ingroup markers.

While our model provides valuable insights into the mechanisms of information integration and consensus formation, several limitations should be acknowledged. First, the conclusions would benefit from empirical validation and a more rigorous mathematical analysis to better understand system stability and convergence. Second, the model currently represents a closed system with homogeneous interaction patterns shaped by singular identities, omitting potential asymmetries and external influences (e.g., media or institutional contexts). Future research should incorporate diverse mechanisms of cross-community information flow, including top-down sources, varying trust parameters and multiple identities, to more accurately capture the complexities of political opinion formation.

\section{Conclusion}

With this model, we investigated a bidirectional mechanism of political opinion dynamics in democratic societies and how it might be affected by changes in inter-group connectivity. We found that hierarchical coarse-graining and community-bounded social influence with a probability of inter-community migration of agents can lead to the emergence of downward causation, revealing the complete upward and downward causal mechanism leading to consensus in our model as well as three regimes of information integration. The iterative regime, with a small degree of higher-level connectivity gives rise to transient diversity and prolonged time until consensus. This allows for slow variable construction which increases the fidelity of the constructed coarse-grained variable to complex and slow-changing systems \cite{Flack2017a, Smaldino2022}. When the migration of agents between communities has a higher probability, information integration across labels happens parallel to local information integration and quickly leads to global alignment before local alignment, preventing a transient diversity from arising.

Polarization can be understood as a society’s inability to coarse-grain to the macro-level with locally diverging perceptions of issues along group lines. In a network like the one here, where information integration occurs within epistemic ingroups, any coherent information integration between communities leads to global consensus via the emergence of global downward causation. An explanation for polarization may therefore involve incoherence of tuning at the higher level that produces alignment within but misalignment across groups. Overall, the discovery of the causal mechanism behind the opinion dynamics in our model, including both upward and downward causation, can further the formulation of research on the causes of, and interventions for dysfunctions of political discourse.

\section{Acknowledgements}

This research paper would not have been possible without the exceptional guidance and support of the BIGSSS Summer School on Democratic Debate from
July 03 - July 12, 2023 organized by Jan Lorenz, Constructor University, Bremen, Germany and Marijn Keijzer, Institute for Advanced Study in Toulouse, University of Toulouse, France.

\bibliographystyle{unsrt}  


\end{document}